# Artificial Neural Network Based Modeling on Unidirectional and Bidirectional Pedestrian Flow at Straight Corridors

**Xuedan Zhao[1], Long Xia[1], Jun Zhang[1,*], Weiguo Song[1]**

[1]State Key Laboratory of Fire Science, University of Science and Technology of China, Jinzhai Road 96, Hefei, Anhui, People's Republic of China

**Abstract**

Pedestrian modeling is a good way to predict pedestrian movement and thus can be used for controlling pedestrian crowds and guiding evacuations in emergencies. In this paper, we propose a pedestrian movement model based on artificial neural network. In the model, the pedestrian velocity vectors are predicted with two sub models, Semicircular Forward Space Based submodel (SFSB-submodel) and Rectangular Forward Space Based submodel (RFSB-submodel), respectively. Both unidirectional and bidirectional pedestrian flow at straight corridors are investigated by comparing the simulation and the corresponding experimental results. It is shown that the pedestrian trajectories and the fundamental diagrams from the model are all consistent with that from experiments. And the typical lane-formation phenomena are observed in bidirectional flow simulation. In addition, to quantitatively evaluate the precision of the prediction, the mean destination error (MDE) and mean trajectory error (MTE) are defined and calculated to be approximately 0.2m and 0.12m in unidirectional flow scenario. In bidirectional flow, relative distance error (RDE) is about 0.15m. The findings indicate that the proposed model is reasonable and capable of simulating the unidirectional and bidirectional pedestrian flow illustrated in this paper.

## 1. Introduction

Accidents resulting from the crowd crushes and stampedes occur frequently causing tremendous loss of life and property. The study on pedestrian dynamics can provide efficient strategies and approaches to guide pedestrian flow and thus reduce accident loss. While pedestrian movement modelling is a primary and effective method to simulate and predict pedestrian movement.

In the past decades, scholars have built hundreds of pedestrian movement models[1]. These models can be divided into the macroscopic models and microscopic models. The former generally consider the relationships among system's characteristics like density, flow and mean speed from the system level rather than a single unit [2]. In contrast to macroscopic models, microscopic models focus on an individual in pedestrian flow to represent his/her microscopic properties like position and velocity. In the term of the continuity of time and space, microscopic models generally fall into two categories, discrete models and continuum models. Cellular Automata (CA) models [3] and Lattice Gas (LG) models [4] are two typical discrete models. Opposite of the discrete models, the time, space and pedestrian motion state in continuum models are all continuous. Force-based model like Social Force Model (SFM) [5] and Centrifugal Force Model (CFM) [6], as a typical continuum models, is based on the assumption that the motion of a pedestrian is driven by the virtual forces from the agent itself, surrounding pedestrians and the walls.

Despite that the current pedestrian movement models have presented many significant phenomena and have a certain capacity to describe pedestrian motion characteristics in some specific scenarios, there still remains some problems and challenges. For instance, the modeling process is relatively complex because of various mathematical formulas and movement rules. It is difficult to take individual differences of pedestrians and other complex influence factors into account. Furthermore, the parameters in most of pedestrian movement models are adjusted using small part of experimental data, which makes the generalization ability of model poor.

Machine learning, as a subset of Artificial Intelligence, is used in a wide variety of applications. In the field of pedestrian dynamics, traditional machine learning algorithms such as Support Vector Regression (SVR)[7], Reinforcement Learning (RL)[8] have been used to improve the modeling of pedestrian dynamics. Moreover, recently artificial neural network is rapidly developing and also wildly used in various research fields owing to its advantages in dealing with nonlinearity, uncertainty and other problems. To overcome the above-mentioned shortages, there appear some works on pedestrian dynamics based on artificial neural network. Shao [9] developed a fundamental model of pedestrian simulation based on Cellular Automata. In order to consider the terrain factors, neural network was embedded into pedestrians ensuring agents motion more intelligent and realistic. Ma *et al.* [10] proposed a novel approach for simulating pedestrian movement behavior based on artificial neural network. The neural network was trained utilizing experimental data collected from a realistic road crossing with bidirectional flow. Simulation results were reasonably consistent with the real-life scenario and acceptable from an engineering perspective. Song *et al.* [11] proposed a multi-scenario adaptive neural network to model pedestrian behavior. Normalization of relative positions among pedestrians and speed direction transfer algorithm were used to train neural network. It is shown that the proposed neural network was capable of reproducing more realistic flow in multiple scenarios. Besides, other newly developed deep learning method, such as generative adversarial network (GAN) [12,13], also has begun applying on pedestrian trajectory prediction. However, the pedestrian movement modeling researches based on artificial neural network are still insufficient and are necessary to make further studies. Motivated by that, a pedestrian movement model based on artificial neural network is proposed in this paper to mimic pedestrian behaviors in both unidirectional and bidirectional flow at straight corridors, which are two typical and common phenomena in our real life. Recently, A lot of works about unidirectional and bidirectional flow have been conducted by many researchers including experiments [14,15] and models [16,17]. Scholars still pay attention on the studies of unidirectional and bidirectional flow due to unknown parts of crowd movement mechanism and especially the low quality of reproduced experiment trajectories given by existing pedestrian models.

The organization of this paper is as follows. Section 2 detailedly illustrates the model structure and implementation details. In Section 3, the unidirectional and bidirectional flow experiment datasets are introduced. Data preprocessing methods and the details of network training are also presented in Section 3. Then in the next section, the model is used to mimic the pedestrian unidirectional and bidirectional flow at straight corridors. The simulation results are analyzed and compared with experiments at the same scenarios to validate the proposed model. Section 5 ends the paper with the main conclusions and some future works.

**Nomenclature**

| | |
|---|---|
| SFSB | Semicircular Forward Space Based |
| RFSB | Rectangular Forward Space Based |
| $(x_0, y_0)$ | Coordinate of the subject pedestrian |
| $(x_j, y_j)$ | Coordinate of the pedestrian $j$ |
| $|\vec{v}_0|$ | Magnitude of subject pedestrian' velocity |
| $\theta_{\vec{v}_0}$ | Direction of subject pedestrian' velocity |
| $\vec{L}_j$ | Forward distance vector from $(x_0, y_0)$ to $(x_j, y_j)$, whose maginitude is $|\vec{L}_j|$ |
| $\vec{I}_j$ | Input parameters vector of network in SFSB-submodel |
| $\vec{I}_{j,x}, \vec{I}_{j,y}$ | $x$, $y$ components of $\vec{I}_j$, whose magnitudes are respectively $|\vec{I}_{j,x}|$, $|\vec{I}_{j,y}|$ |
| $\theta$ | Direction angle of $\vec{L}_j$, $\in (-\pi, \pi]$ |
| $n_p$ | The number of experimental pedestrians in semicircular forward space |
| $n_w$ | The number of discrete wall pieces in semicircular forward space |
| $n_{c,i}$ | The number of total objects (pedestrians and discrete wall pieces) in semicircular forward space for Network $i$ |
| $N_{in}$ | The number of neurons in input layer |
| $N_{out}$ | The number of neurons in output layer |
| $N_{h1}$, $N_{h2}$ | The number of neurons in first and second hidden layer |
| $b_{ent}$ | Width of the corridor entrance (*m*) |
| $b_{cor}$ | Width of the corridor (*m*) |
| $b_{exit}$ | Width of the corridor exit (*m*) |
| $E$ | Mean square error |
| $O$ | Output of neural network |
| $T$ | Target value |
| $MDE$ | Mean destination error |
| $DE_i$ | Destination error of pedestrian $i$ |
| $(x_i^D, y_i^D)_{exp}$ | Coordinate of pedestrian at the last time step in experiment |
| $(x_i^D, y_i^D)_{sim}$ | Coordinate of pedestrian at the last time step in simulation |
| $MTE$ | Mean trajectory error (*m*) |
| $\overline{MTE}$ | The mean of $MTE$ (*m*) |
| $TE_t$ | Location error of one pedestrian in simulation and experiment at time step $t$ (*m*) |
| $totalTS$ | The number of total time steps |
| $(x_t, y_t)_{exp}$ | Coordinate of pedestrian at time $t$ in experiment |
| $(x_t, y_t)_{sim}$ | Coordinate of pedestrian at time $t$ in simulation |
| $J$ | Pedestrian flow (*ped/s*) |
| $\rho$ | Pedestrian density (*ped/m²*) |
| $v$ | Pedestrian movement speed (*m/s*) |

**2. Model**

Before introducing the model, we firstly make the following definitions and assumptions.

a. The studied pedestrian at current moment is called as subject pedestrian.
b. As shown in Fig. 1, the forward space of subject pedestrian is a semi-infinite space, whose boundary passes through the coordinate points of the subject pedestrian and is perpendicular to the movement direction of the system consisting of overall pedestrians.
c. Only the pedestrians in the forward space affect the movement of the subject pedestrian. The pedestrians at the rear have little influence on the movement of the subject pedestrian.
d. To reduce the calculation amount, we choose 2.1m as the standard distance to process the forward space in the rest of the article, since the pedestrians 2.1m away have little impact on the subject pedestrian's movement [18].

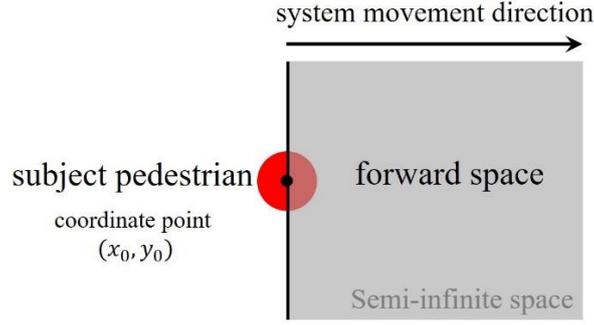

Fig. 1 The sketch of the subject pedestrian's forward space.

The overall framework of pedestrian movement model proposed in this paper is shown in Fig. 2. According to different requirements, the forward space of the subject pedestrian is processed into two shapes, semicircle and rectangle. The model consists of two sub models, which are respectively built based on the semicircular forward space and the rectangular forward space. The Semicircular Forward Space Based sub model (i.e. SFSB-submodel) mainly utilizes the data produced from the forward distance to obtain the magnitude of pedestrian's current velocity. The direction of velocity is derived from the Rectangular Forward Space Based sub model (i.e. RFSB-submodel) according to the pedestrian distribution. The detail information about the both of submodels will be introduced in the next two subsection.

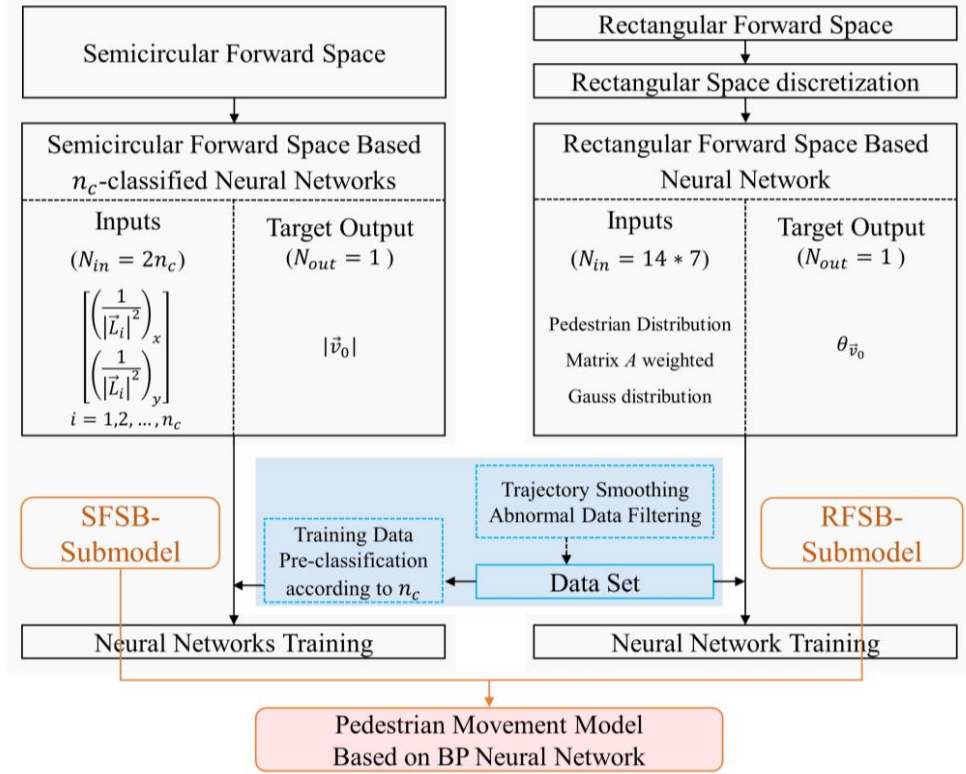

Fig. 2 System flowchart. The pedestrian movement model based on BP neural network consists of two sub models, the Semicircular Forward Space Based sub model (SFSB-Submodel) and the Rectangular Forward Space Based sub model (RFSB-submodel).

It should be noted that in the building process of our model, several attempts of model structure were conducted to find out the best fit for our model assumption and neural network structure. These attempts are list in Table 1. After simply comparing the results of these models, it was found that the forth model performed better than the others did. Model building trials showed that the combination of discrete and continuous perceptive region is a promising research area. Therefore, in the paper, we only introduce the third model consisting of two sub models in detail.

Table 1. Attempts of different models.

| Model Index | Sub-model | Shape of forward space | Continuity | Output of network |
|---|---|---|---|---|
| 1 | Only SFSB | semicircular | continuous | magnitude and direction of velocity |
| 2 | Only RFSB | rectangular | discrete | magnitude and direction of velocity |
| 3 | SFSB | semicircular | continuous | direction of velocity |
|   | RFSB | rectangular | discrete | magnitude of velocity |
| 4 | SFSB | semicircular | continuous | magnitude of velocity |
|   | RFSB | rectangular | discrete | direction of velocity |

2.1 Semicircular Forward Space Based submodel (SFSB-submodel)

The semicircular forward space based sub model are mainly used to predict the magnitude of pedestrians' velocity. Here, we choose a three-layer BP neural network as the basic framework of this sub model. The output of the network is the magnitude of pedestrians' velocity at current moment. The selection of input parameters is based on the forward distance of pedestrians, which includes many useful information about pedestrian behavior. Next, we will elaborate the whole process of the input parameters selection.

Although the effect on subject pedestrian is difficult to quantify, fortunately, the forward distance can reflect the interaction between the subject pedestrian and other pedestrians in a degree. So, in the term of the forward distance vector, we choose $\vec{l}_j$ as the input parameters of network. The magnitude of $\vec{I}_j$ is $1/|\vec{L}_j|^2$, and the direction is as same as that of the forward distance vector $\vec{L}_j$. So, the $x$, $y$ components $\vec{I}_{j,x}, \vec{I}_{j,y}$ of $\vec{I}_j$ can be calculated with formulas below:

$$|\vec{I}_{j,x}| = \frac{1}{|\vec{L}_j|^2} \cdot cos\theta \quad (1)$$

$$|\vec{I}_{j,y}| = \frac{1}{|\vec{L}_j|^2} \cdot sin\theta \quad (2)$$

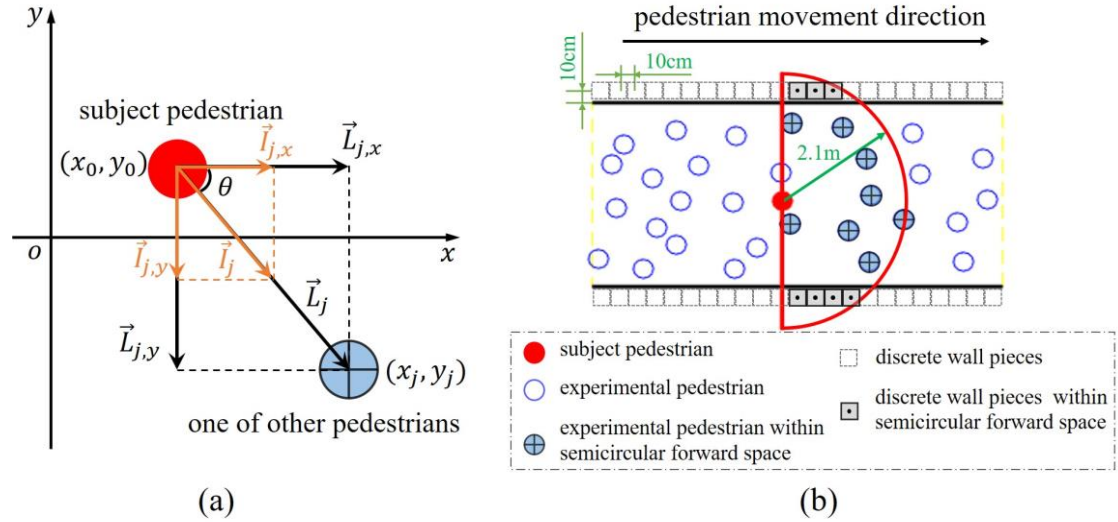

Fig. 3 The sketch of semicircular forward space of subject pedestrian. (a) The vector of forward distance $\vec{L}_j$ from the subject pedestrian to one of the other pedestrians with two corresponding components $\vec{L}_{j,x}, \vec{L}_{j,y}$ along $x$ and $y$ axes. (b) The semicircular forward space of the subject pedestrian (red solid circle) with a radius of 2.1m includes some experimental pedestrians (blue solid circle with +) and some discrete wall pieces (grey solid rectangle with •).

According to the assumption d, the forward space of the subject pedestrian in this section is a red solid semicircle with 2.1m radius, as shown in Fig. 3(b). We choose the red solid circle as the subject pedestrian for example. In his/her semicircular forward space, there are some experimental pedestrians (blue solid circle with +). These experimental pedestrians have an effect on the movement of the subject pedestrian, so do the walls. In order to considering the effect of the walls, we divided the continuous walls into many small discrete wall pieces, which is similar with the discretization idea of cellular automata model [3]. These small discrete wall pieces are arranged (grey dotted hollow rectangles) at interval of 10*cm* along the lines extending outward for 10*cm* from both walls. Now the objects which can affect the motion the subject pedestrian include the discrete wall pieces and experimental pedestrians within the semicircular forward space.

$$n_c = n_p + n_w \qquad (3)$$

Where $n_p$, $n_w$ and $n_c$ respectively denote the number of experimental pedestrians, discrete wall pieces and the total objects in the semicircular forward space.

The $n_c$ has different values for different subject pedestrian, which means the sample sets we produced will have different dimensions. So, we can't use these sample sets to train a network. To solve this problem, we build many networks. The number of networks keeps consistent with that of $n_c$ values. The architecture of the $i$-th neural network in this section which has three layers is shown in Fig. 4. According to the total number of pedestrian in semicircular forward space $n_{c,i}$, the input parameters is

$$\{|\vec{I}_{j,x}|, |\vec{I}_{j,y}|, \quad j = 1, 2, \ldots, n_{c,i}\} \qquad (4)$$

The magnitude of the subject pedestrian' velocity $|\vec{v}_0|$ is as the target output. For hidden layers, how many neurons should be used in hidden layers is still a problem[19]. So, we made an attempt according to some "rules of thumb"[19] for choosing the hidden layer units. The number of neurons in hidden layer is given by

$$N_h = \left\lceil \frac{N_{in} + N_{out}}{2} \right\rceil \qquad (5)$$

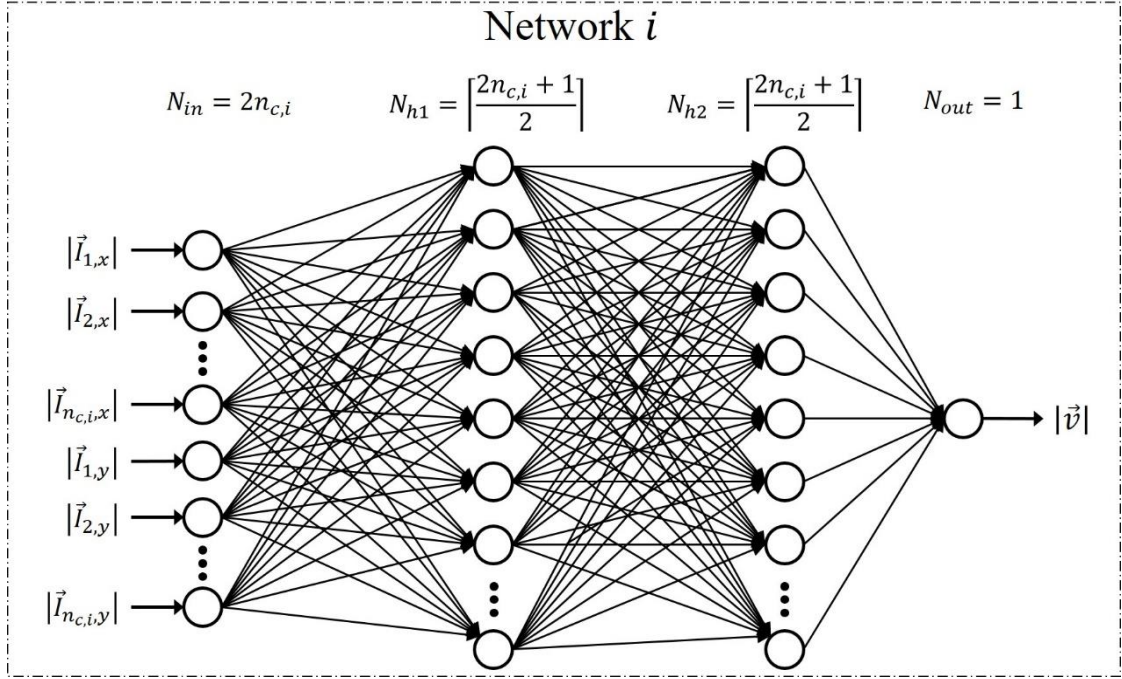

Fig. 4 The architecture of the neural network in semicircular forward space based submodel, including the input and output parameters and the number of neurons in per layer.

## 2.2 Rectangular Forward Space Based submodel (RFSB-submodel)

The rectangular forward space based sub model are mainly used to predict the direction of pedestrians' velocity. As same as the semicircular forward space based sub model, we also use a three-layer BP neural network as the basic framework of rectangular forward space based sub model. The output of this network is the direction of pedestrians' velocity at current moment. The input parameters are selected as follows.

Pedestrian spatial distribution is a representation of pedestrian movement and can affect the motion of pedestrians. Therefore, we choose the pedestrian spatial distribution as input parameters of the network. To quantify the pedestrian spatial distribution, we firstly chose a 4.2 *m*×2.1 *m* (selected according to assumption d) rectangle as the forward space of the subject pedestrian, as shown in Fig. 5. Then, the rectangle was subdivided into many square grids. In order to ensure pedestrians occupy a same grid as little as possible, after several trials, 0.3 m×0.3 m grid size was chosen. Now we can use a two-dimensional matrix with seven rows and fourteen columns to represent the pedestrian distribution. The value of the element in the matrix corresponding to a grid will be changed from zero to one, when there is a pedestrian locating in the grid. The rest of elements' values still keep zero. We also considered the effect of walls. If the center of a grid is outside of either wall, then the value of the element in the matrix corresponding to the grid will be also changed to one. Besides, the pedestrians who is closer to the subject pedestrian have a stronger effect on the movement of the subject pedestrian. To consider this effect, the input parameters of the network is the pedestrian spatial distribution weighted the gauss distribution, as shown in Fig. 5.

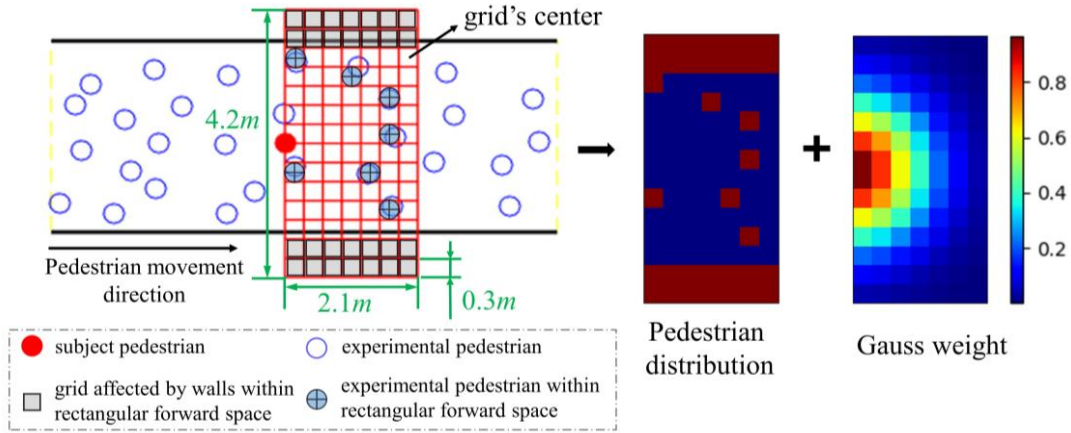

Fig. 5 The sketch of rectangular forward space of the subject pedestrian and the corresponding pedestrian spatial distribution.

Fig. 6 gives the architecture of the network in this section. The pedestrian spatial distribution matrix has been changed to one dimension, so the number of the input neurons is 7×14=98. The same as Section 2.1, as an attempt, the number of neurons in the first hidden layer is 50 calculated by Equation (5). And the number in the second hidden layer is 25, a half of that in the first hidden layer. The direction of pedestrians' velocity $\theta_{\vec{v}_0}$ at the current moment is choose as the target output of the network.

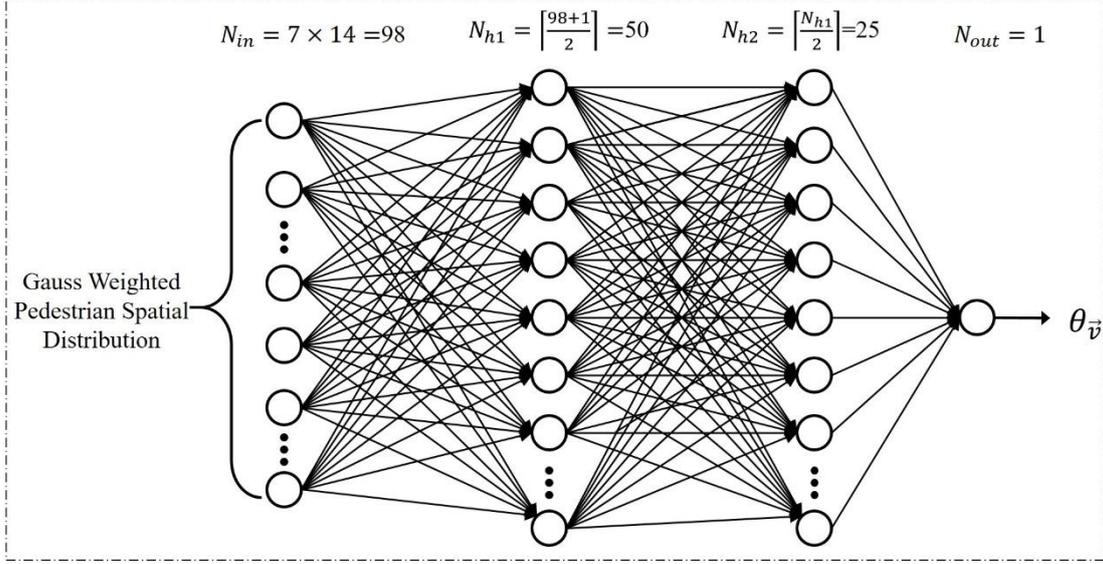

Fig. 6 The architecture of the neural network in rectangular forward space based submodel, including the input and output parameters and the number of neurons in per layer.

2.3 Implementation Details

The Backpropagation algorithm is used to train the networks. Errors between network outputs and target outputs are calculated based on the current weights of the network. Then stochastic gradient descent is used to adjust and optimize the weights to reduce the errors of the network. We repeat this procedure until the training errors are reduced to an acceptable scope. Learning rate is fixed at 0.01. The activation function in hidden and output layers are respectively tan-sigmoid and rectified linear unit. In order to calculate the error between output values of the network and the target values, we use the mean square error as the loss function shown below

$$E = \sqrt{\frac{1}{N}\sum_{i=1}^{N}|O_i - T_i|^2} \qquad (6)$$

Where, $O_i$ is the output of network, $T_i$ is the corresponding target value, $N$ is the total number of training samples.

In order to prevent the training overfitting, we applied the early-stop validation approach to supervise the training. When the training samples are input into the network, they will be automatically and randomly divided into training sets, validation sets and testing sets according to a certain proportion (80%, 15% and 5% in this paper). These three data sets are independent with each other. The validation sets are mainly used to detect overfitting during training. When an iteration is completed, the validation sets will be applied on the current-state trained network to obtain validation errors.

## 3. Experiments

In the sections of Experiments and Results, we evaluate our model on two scenarios: unidirectional flow scenario and bidirectional flow scenario. The simulation results are compared with that of experiments and other's works from both qualitative and quantitative perspectives.

3.1 Datasets

3.1.1 Unidirectional flow

Unidirectional flow data are from a dataset built up by the institute section Civil Safety Research in the Research Centre Jülich in Germany, which is a professional pedestrian dynamical dataset under laboratory experimental condition and publicly available on [20]. The sketch of experimental setup and two snapshots are shown in Fig. 7. The length and width of the corridor are 8m and 3m. The experiments were conducted with up to 350 participants who were composed mostly of students. The experiments were recorded by two cameras (frame rate: 16 fps) mounted at the rack of the ceiling of the hall. The pedestrian trajectories were automatically extracted from the video recordings by using the software *PeTrack* [21] and were directly downloaded from the website http://ped.fz-juelich.de/da/2009unidirOpen. The detailed information on the data extraction principle of the software can be found in [30]. The free speed 1.55± 0.18 m/s was obtained by measuring 42 participants' free movement [22]. We choose six runs trajectories of the experiments with different pedestrian densities to generate the training samples, as shown in Table 2.

Table 2. Parameters of unidirectional flow experiments in the straight corridor

| Experiment index | Name | $b_{ent}/m$ | $b_{cor}/m$ | $b_{exit}/m$ | $N$ |
|---|---|---|---|---|---|
| 1 | UF-080-300-300 | 0.80 | 3.00 | 3.00 | 119 |
| 2 | UF-100-300-300 | 1.00 | 3.00 | 3.00 | 100 |
| 3 | UF-120-300-300 | 1.20 | 3.00 | 3.00 | 163 |
| 4 | UF-180-300-300 | 1.80 | 3.00 | 3.00 | 208 |
| 5 | UF-240-300-300 | 2.40 | 3.00 | 3.00 | 296 |
| 6 | UF-300-300-300 | 3.00 | 3.00 | 3.00 | 349 |

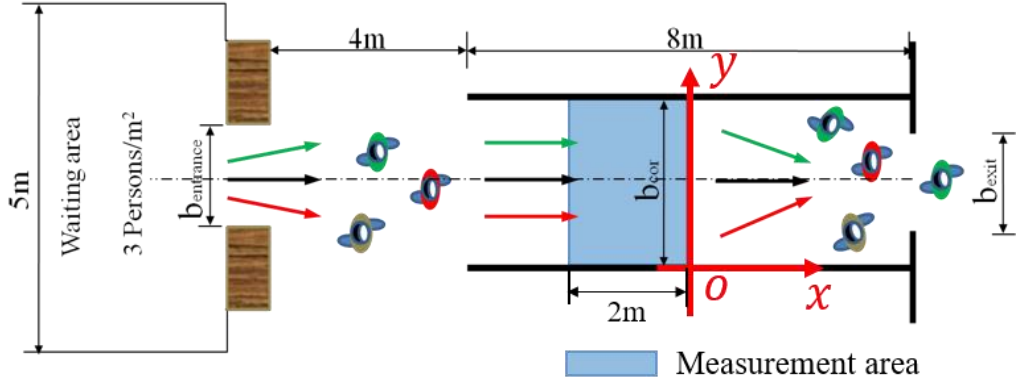

(a)

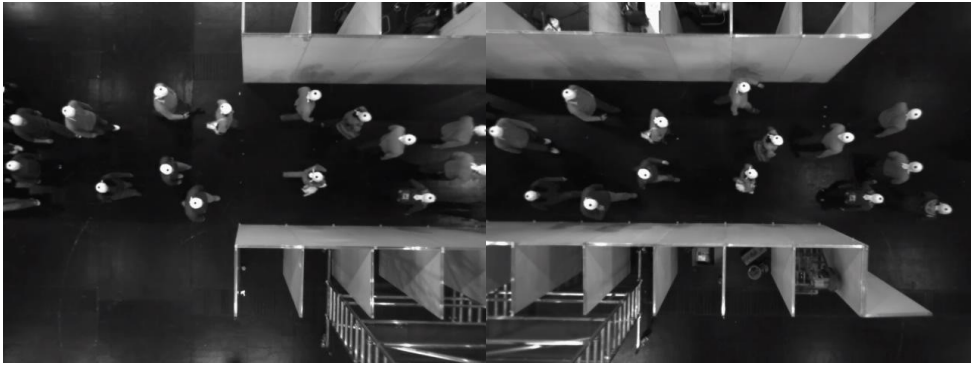

(b)

Fig. 7 (a) Sketch of the unidirectional flow experimental setup [22]. (b) Snapshots of the experiment UF-100-300-300 from two cameras respectively.

3.1.2 Bidirectional flow

Bidirectional flow data are also from the dataset built up by the institute of Civil Safety Research in the Research Centre Jülich in Germany [20]. The experimental setup of bidirectional flow is highly consistent with that of the unidirectional flow described in subsection 3.1.1. The sketch of the experiment setup is shown in Fig. 8. The length and width of the corridor are 8m and 3m. The experiments were conducted with up to 350 participants who were composed mostly of students. The experiments were recorded by two cameras (frame rate: 16 fps) mounted at the rack of the ceiling of the hall. The pedestrian trajectories were directly downloaded from the website http://ped.fz-juelich.de/da/2009bidirSym. The average free speed 1.55± 0.18 m/s was obtained by measuring 42 participants' free movement [23]. We choose five runs trajectories of the experiments with different pedestrian densities to generate the training samples, as shown in Table 3.

Table 3. Parameters of bidirectional flow experiments in the straight corridor

| Experiment index | Name | $b_l$/m | $b_{cor}$/m | $b_r$/m | $N_l$ | $N_r$ |
|---|---|---|---|---|---|---|
| 1 | BF-050-300-050 | 0.50 | 3.00 | 0.50 | 54 | 71 |
| 2 | BF-065-300-065 | 0.65 | 3.00 | 0.65 | 64 | 83 |
| 3 | BF-075-300-075 | 0.75 | 3.00 | 0.75 | 61 | 86 |
| 4 | BF-085-300-085 | 0.85 | 3.00 | 0.85 | 119 | 97 |
| 5 | BF-100-300-100 | 1.00 | 3.00 | 1.00 | 125 | 105 |

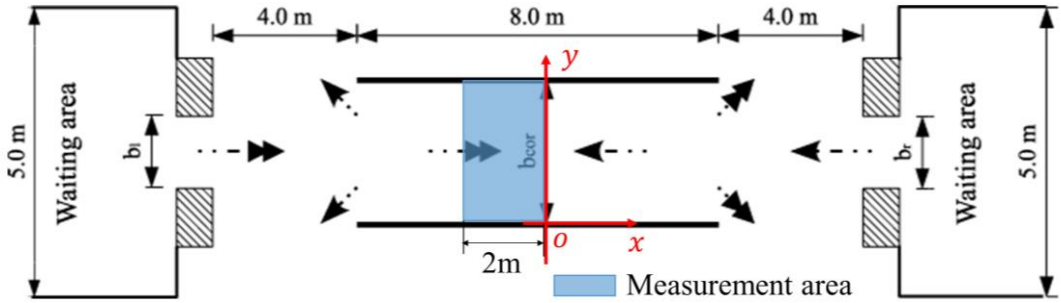

Fig. 8 Sketch of setup in the bidirectional flow experiment [23].

3.2 Data Preprocessing

Before generating the training samples, we preprocessed these pedestrian trajectories. First, we smoothed the pedestrian trajectories to avoid the effect of pedestrian body swaying. The direction of pedestrian bodies swaying is the y direction, so we only processed the y coordination. We use the mean filter, which means the y coordination at time $t$ is replaced by the mean of that at time $t-1$ and time $t+1$. Then, cubic spline interpolation was applied on the midpoints obtained between peak points and neighboring trough points. Fig. 9 gives the initial and smoothed pedestrian movement trajectories in two different density. Other trajectory smoothing methods [24,25] are suitable for this preprocess as well. Second, PauTa criterion method [26] was applied on the data points consisting of the magnitude and direction of velocities of all pedestrians at each moment to remove the data including gross error. The red bold lines shown in Fig. 10 are the results. These data were removed when generating training samples.

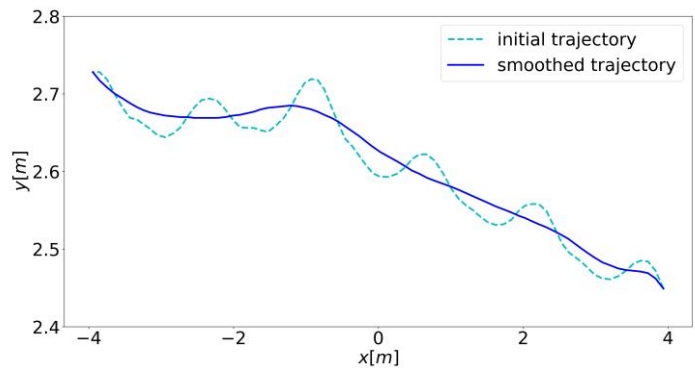

(a)

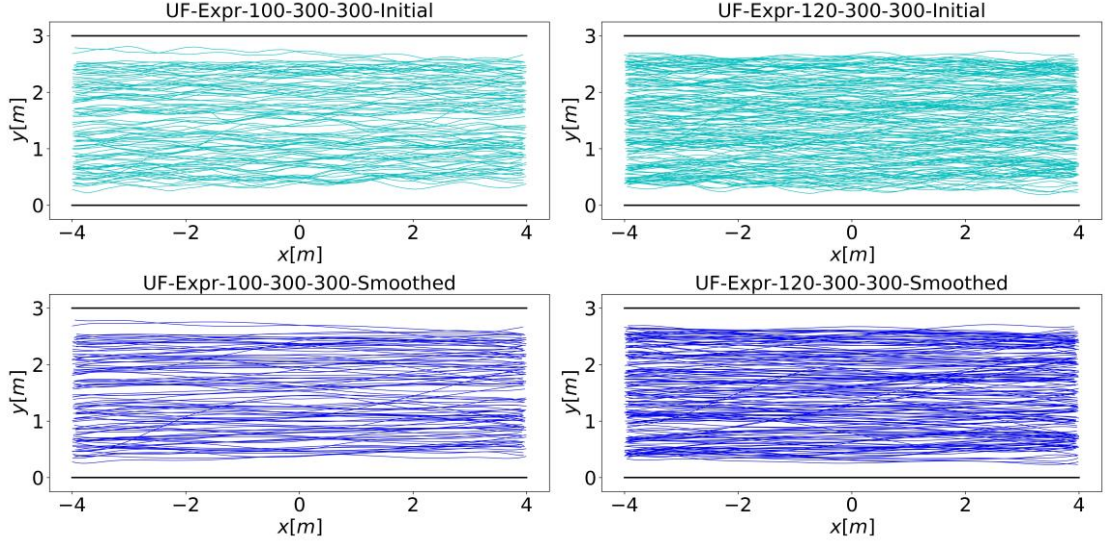

(b)

Fig. 9 (a) An individual's initial experimental trajectory and corresponding smoothed trajectory. (b) All pedestrians' initial experimental pedestrian trajectories and corresponding smoothed trajectories under two different densities.

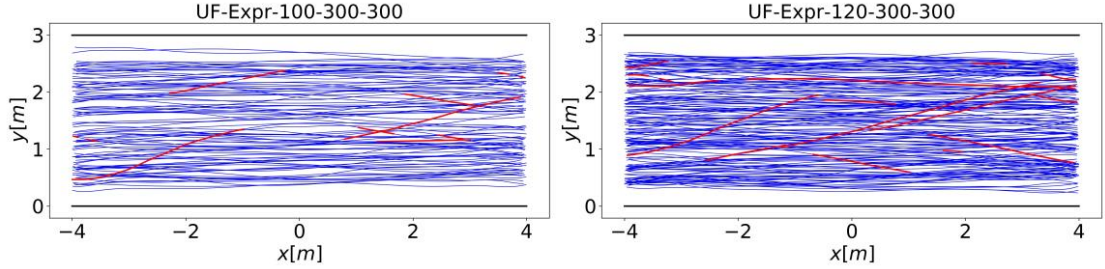

Fig. 10 The initial experimental pedestrian trajectories and the detected abnormal movement data (red lines) using pauta criterion method.

3.3 Network Training

For preprocessed pedestrian trajectories in unidirectional flow, we chose a time step of one frame (16 frames = 1s) to generate training samples. The number of training samples of each network in unidirectional flow scenario and bidirectional flow scenario are listed in Table 4 and Table 5 respectively. Especially, in bidirectional flow scenario, according to the different pedestrian movement direction (from left to right and from right to left), the samples are divided into two groups to respectively train the networks for the better performance of the model.

Table 4. Number of training samples in unidirectional flow scenario.

| Submodel | $n_{c,i}$ | Network Architecture | | Number |
|---|---|---|---|---|
| | | Network $i$ | $N_{in} \times N_{h1} \times N_{h2} \times N_{out}$ | |

| Submodel | $n_{c,i}$ | Network $i$ | $N_{in} \times N_{h1} \times N_{h2} \times N_{out}$ | Number |
|---|---|---|---|---|
| | 1 | 1 | 2×2×2×1 | 874 |
| | 2 | 2 | 4×3×3×1 | 1355 |
| | … | … | … | … |
| SFSB | 27 | 27 | 54×28×28×1 | 12014 |
| | … | … | … | … |
| | 41 | 41 | 82×42×42×1 | 90 |
| | 42 | 42 | 84×43×43×1 | 13 |
| RFSB | | | 98×50×25×1 | 131219 |

Table 5. Number of training samples in bidirectional flow scenario.

| Submodel | Network Architecture | | | Number | |
|---|---|---|---|---|---|
| | $n_{c,i}$ | Network $i$ | $N_{in} \times N_{h1} \times N_{h2} \times N_{out}$ | L→R | R→L |
| | 1 | 1 | 2×2×2×1 | 567 | 490 |
| | 2 | 2 | 4×3×3×1 | 502 | 483 |
| | … | … | … | … | … |
| SFSB | 27 | 27 | 54×28×28×1 | 3696 | 3097 |
| | … | … | … | … | … |
| | 41 | 41 | 82×42×42×1 | 4 | 12 |
| | 42 | 42 | 84×43×43×1 | 2 | 2 |
| RFSB | | | 98×50×25×1 | 51424 | 50173 |

Here, we take the unidirectional flow scenario as an example to describe the training process and the convergence of the networks. Fig. 11 presents the training process of network and the performance of the trained networks on validation sets and testing sets in unidirectional flow scenario. From Fig. 11(a) and (b), we can see that the mean squared error decreased rapidly in the beginning several epochs, then the decrease rate becomes low. Finally, the error of network converges on a small value. Besides, Fig. 11(c) presents the errors of testing sets applied on the total trained networks in SFSB and RFSB. The errors are almost below 0.15m, which indicates that the trained networks perform well on testing sets.

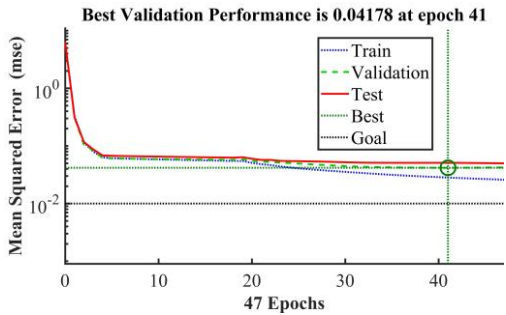 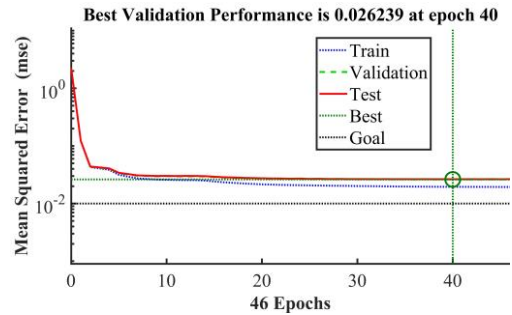

(a)          (b)

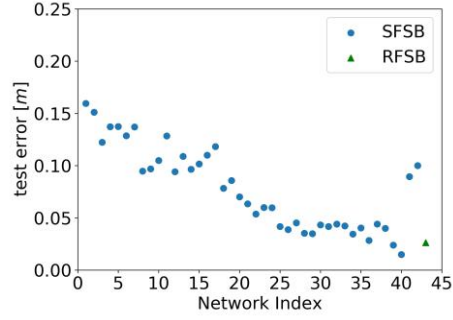

(c)

Fig. 11 The training process of network and the performance of networks on validation sets and testing sets in unidirectional flow scenario. (a) Error profile of 25th network training in SFSB. (b) Error profile of network in RFSB. (c) The performance of trained networks on testing sets.

## 4. Results and Discussions

We apply the proposed model on unidirectional flow and bidirectional flow scenarios to simulate the pedestrian behavior. The dimension of the simulation scenario, the initial location and entering time of each pedestrian in the simulation region are all kept the same as that in the experiments. Then, the movement velocity of each pedestrian at every future time all use the trained networks to predict. The magnitude of velocity uses the SFSB-submodel to predict and the direction of velocity uses the RFSB-submodel to obtain. When using the SFSB-submodel, it must be noted that the number of pedestrians within the semicircular forward space of the subject pedestrian may be more than the maximum of $n_c$ during the simulation. In this case, we can't use the SFSB-submodel to obtain the magnitude of velocity. To address this problem, we give a solution that the pedestrians farthest away from the subject pedestrian are eliminated until the number of rest pedestrians is equal to the maximum of $n_c$. Besides, if no pedestrians are in the semicircular forward space, the magnitude of the velocity is set as 1.55m/s (the average free speed of experiment pedestrians). The simulation time step is 1/16s (one frame) and the parallel update rule is used to update the pedestrians' locations. In addition, at each iteration of the position update, the pedestrians' coordinates are updated in a random sequence. The simulation results are presented and analyzed below.

4.1 Unidirectional Flow Scenario

In the unidirectional flow scenario, the first four experiments with different densities listed in Table 2 are simulated. The simulated pedestrian movement trajectories are presented in Fig. 12. From the figure, we can see that the pedestrian trajectories in the simulations are all basically consistent with the corresponding experiments.

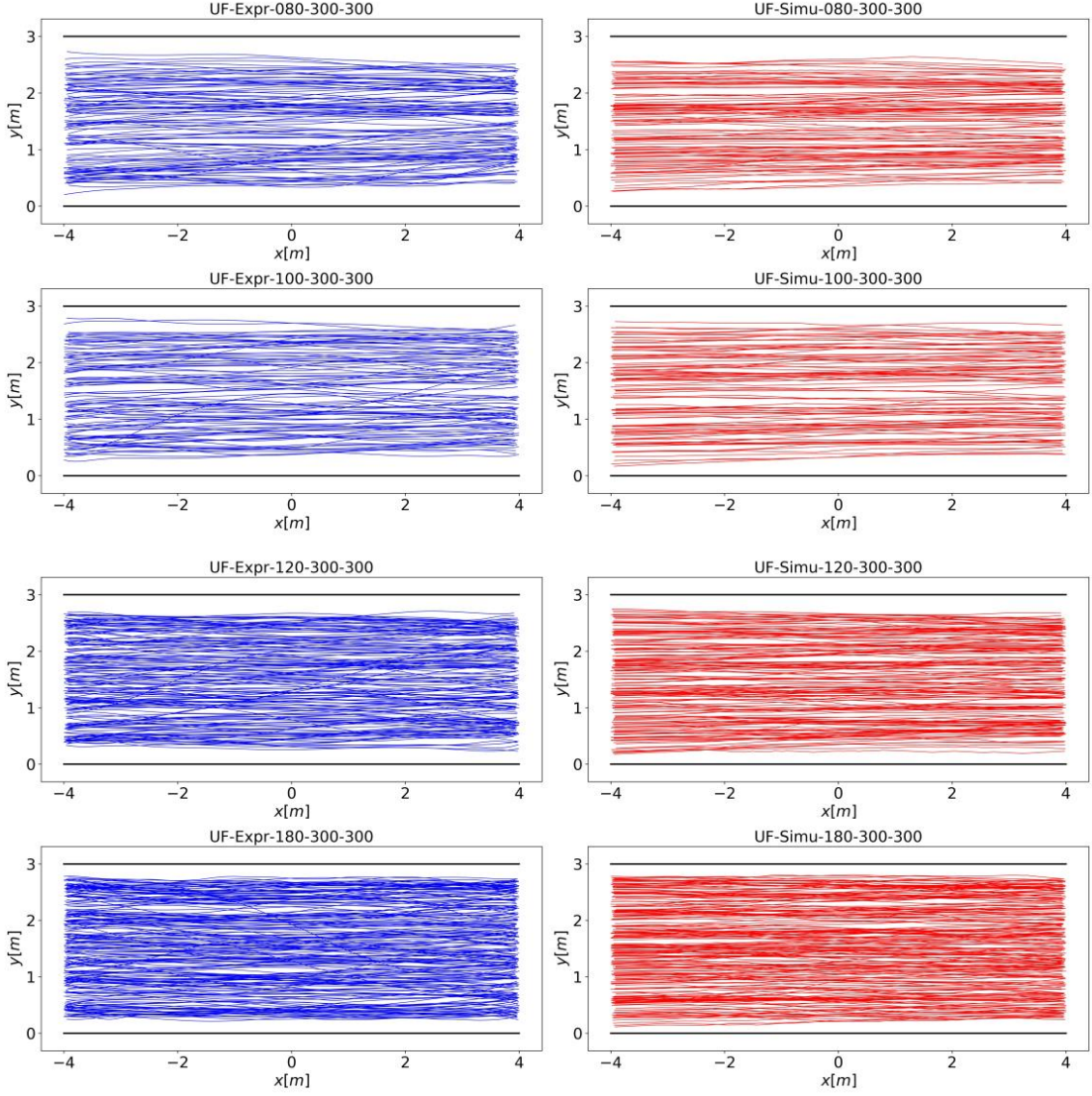

Fig. 12 Comparison of pedestrian trajectories in simulations and experiments with four different densities in unidirectional flow scenario. The pedestrian trajectories in the simulations were all basically consistent with the corresponding experiments.

Moreover, in order to quantitatively illustrate the trajectories error between simulations and experiments, we defined two evaluation indicators, mean destination error (MDE) and mean trajectory error (MTE). The mean destination error is calculated by the formula

$$MDE = \frac{1}{N}\sum_{i=1}^{N} DE_i = \frac{1}{N}\sum_{i=1}^{N}\left[(x_i^D, y_i^D)_{exp} - (x_i^D, y_i^D)_{sim}\right] \quad (7)$$

Where, N is the number of pedestrians in an experiment. $(x_i^D, y_i^D)_{exp}$ and $(x_i^D, y_i^D)_{sim}$ respectively donate the coordinate of the *i-th* pedestrian at the last time step before leaving out of the corridor in the experiment and simulation. $DE_i$ is the distance between $(x_i^D, y_i^D)_{exp}$ and $(x_i^D, y_i^D)_{sim}$. The smaller the MDE is, the smaller the destination error between the simulation and experiment is. Fig. 13 gives the DE-frequency histograms in each case. As shown in this figure, more than 90% pedestrians' destination error (DE) are less than 0.5m and the MDE is all about 0.2m, 2.5% of the corridor length.

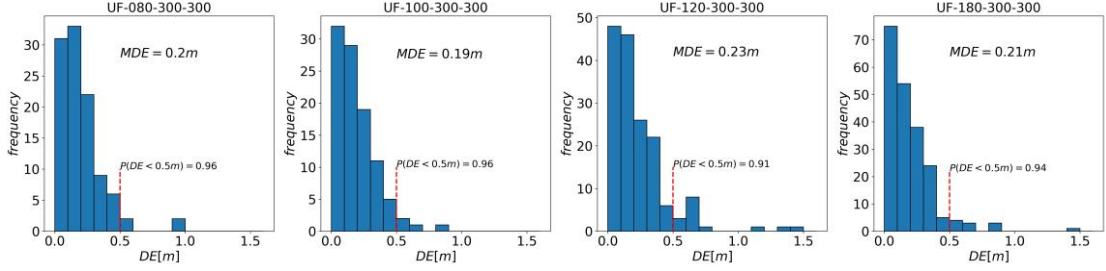

Fig. 13 The DE-frequency histograms of four different cases in unidirectional flow scenario. More than 90% pedestrians' destination error (DE) are less than 0.5m and the MDE are all about 0.2m, 2.5% of the corridor length.

The mean trajectory error of a pedestrian is defined by

$$MTE = \frac{1}{totalTS}\sum_{t=1}^{totalTS} TE_t = \frac{1}{totalTS}\sum_{t=1}^{totalTS}[(x_t, y_t)_{exp} - (x_t, y_t)_{sim}] \quad (8)$$

Where, $TE_t$ represents the location error (distance) of one pedestrian in simulation and experiment at time step $t$. $MTE$ is the mean of $TE_t$ at total time steps ($totalTS$). Fig. 14 gives the MTE-frequency histograms in each case. As shown in this figure, under four different densities, almost all pedestrians' mean trajectory error (MTE) are less than 0.5m. We also calculate the mean of MTE ($\overline{MTE}$)

$$\overline{MTE} = \frac{1}{N}\sum_{i=1}^{N} MTE \quad (9)$$

Which are all about 0.12m, 1.5% of the corridor length. The small mean destination error and mean trajectory error both indicate that the simulation using the model proposed in Section 2 is acceptable in the term of pedestrian movement trajectories.

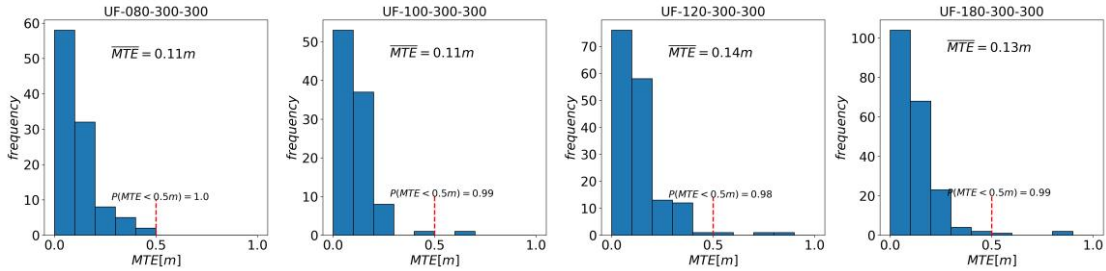

Fig. 14 The MTE-frequency histograms of four different cases in unidirectional flow scenario. Almost all pedestrians' mean trajectory error (MTE) are less than 0.5m and $\overline{MTE}$ are all about 0.12m, 1.5% of the corridor length.

Besides, we also compare the speed-density and flow-density fundamental diagrams between simulations and experiments. The density and speed data both in simulations and experiments are extracted in a 2m×3m rectangle measurement area shown in Fig. 7(a). The density and speed are calculated using the Method B in [22]. Then, the flow is obtained by

$$J = \rho \cdot v \cdot b_{cor} = \rho \cdot v \cdot 3 \qquad (10)$$

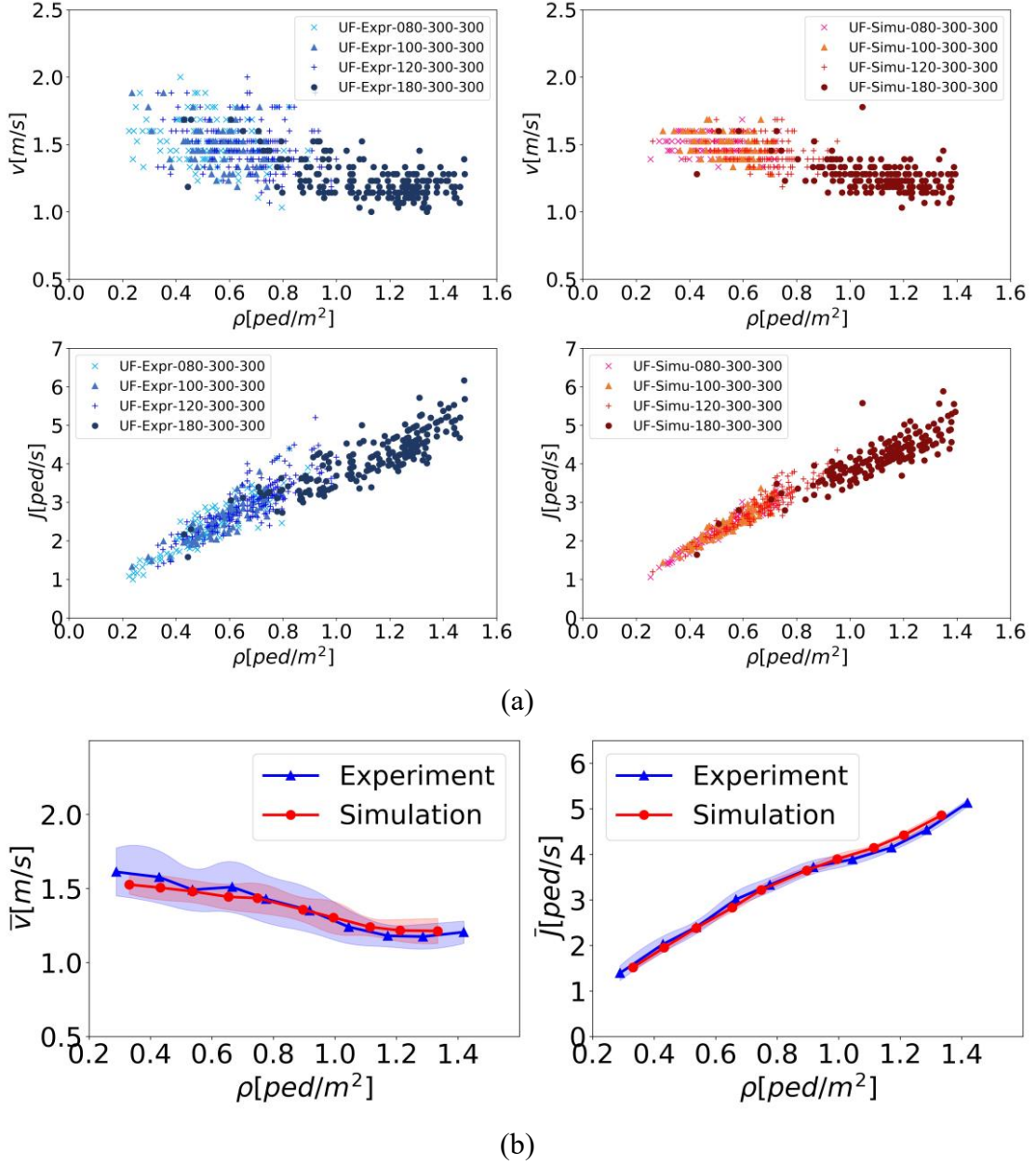

(a)

(b)

Fig. 15 Comparison of fundatmental diagrams between simulations and experiments in unidirectional flow scenario. (a) Density-speed and density-flow diagrams. (b) The mean speed and flow related to density.

The results of four cases in simulations and experiments are shown in Fig. 15(a). The overall distributions of the data points of the simulations are basically in line with the corresponding experiments. In addition, we calculated the mean speed and flow of four cases. Fig. 15(b) shows the results, which also shows that the speed-density and flow-density fundamental diagrams in simulations are basically consistent with the experiments. These results imply that the proposed model in this paper performs well on simulating the pedestrian unidirectional flow illustrated in this paper.

4.2 Bidirectional Flow Scenario

Apart from the unidirectional flow scenario, we also apply our model on bidirectional flow scenario. We simulate the four sets of bidirectional flow experiments with different densities listed in Table 3. Taking the experiment BF-085-300-085 as an example, the pedestrian distributions and trajectories of simulation are showed and compared with that of experiment. From Fig 16 (b), we can see the basic coherence of pedestrian trajectories in simulation and experiment. Moreover, the lane formation phenomena can be obviously observed from Fig 16(a), which indicates our model can reproduce the typical self-organized behavior of pedestrians.

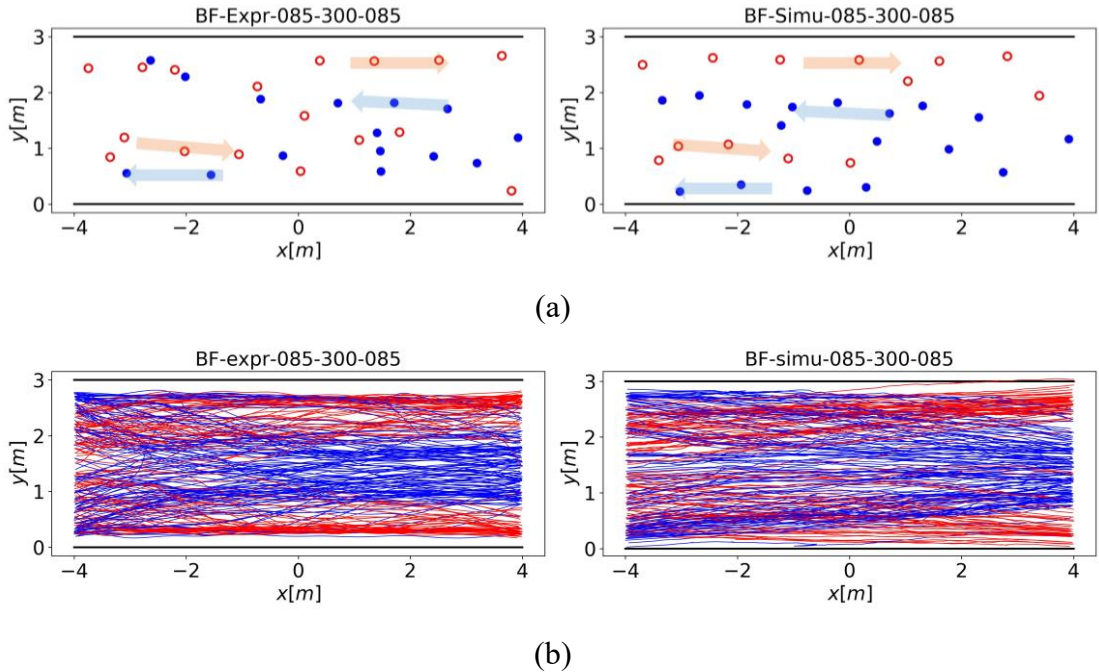

Fig.16 Comparison of pedestrian distributions and trajectories in bidirectional flow scenario. (a) Pedestrian distributions. (b) Pedestrian trajectories, in which the blue and red lines represent pedestrian moving to the left and right, respectively.

In addition, the comparisons of fundamental diagrams between simulation and experiment are presented in Fig.17. The density and speed data are calculated in the measurement area given in Fig. 8. It can be seen from the Fig. 17 that the overall distributions of the data points of the simulations are basically in line with the corresponding experiments.

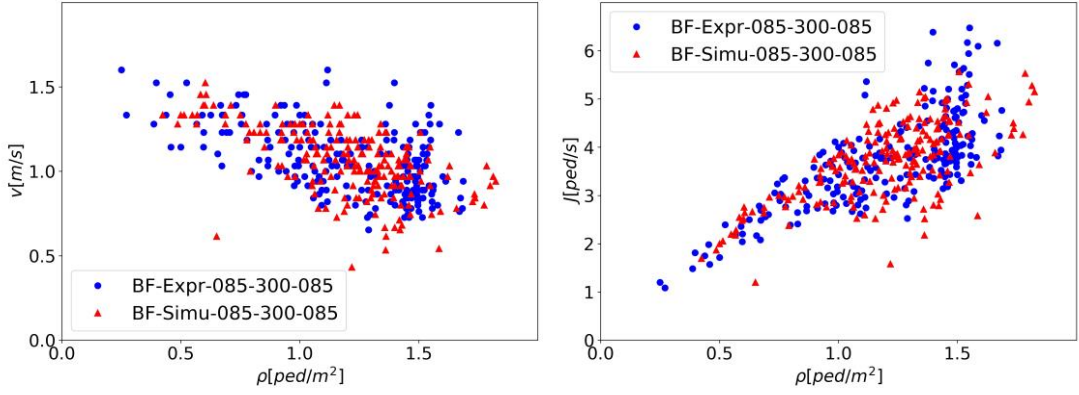

Fig. 17 Comparison of density-speed and density-flow fundamental diagrams between simulation and experiment in BF-085-300-085.

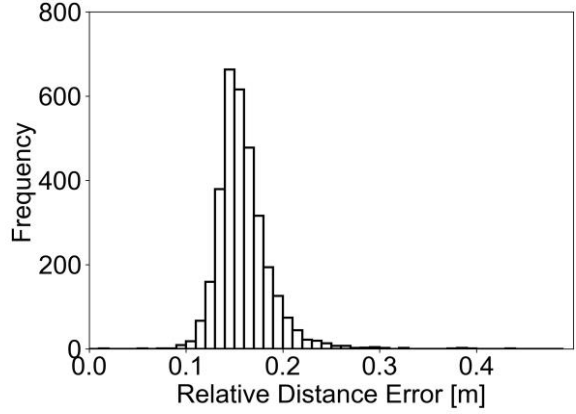

Fig. 18 Frequency distribution of the relative distance error in BF-080-300-085.

To better illustrate the reliability of the proposed model, we also make a quantitative comparison of trajectories. The relative distance error (RDE) [10] at time step $t$ for pedestrian $n$ is defined by

$$E(t,n) = \frac{\|P^{simu}(t+\Delta t,n) - P^{expr}(t+\Delta t,n)\|}{P^{expr}(t+\Delta t,n) - P^{expr}(t,n)} \quad (11)$$

Given $\Delta t = 0.5s$, the relative distance error is calculated over each time step for each pedestrian. Fig. 18 gives the frequency distribution of the relative distance error in BF-080-300-085. The average relative distance errors in four different densities experiments are respectively 0.146m, 0.150m, 0.152m, 0.159m and 0.157m, which are almost a half of the result (0.332m) in ref [10]. These results indicates the proposed model in our paper has a good performance on simulating pedestrian behavior in bidirectional flow scenarios.

## 5. Conclusion

In this paper, a pedestrian movement model based on artificial neural network is proposed. This model consists of two sub models, which are respectively used to learn the magnitude and direction of pedestrian movement velocity. On the basis of this goal, we process the forward space of the subject pedestrian into two shapes, semicircle (radius 2.1m) and rectangle (length×width=4.2m×2.1m). We consider that only the pedestrians within the semicircular and rectangular forward space have an effect on the movement of the subject pedestrian. Besides, the effect of wall is replaced by the visual pedestrians set in a certain way. In the semicircular forward space based submodel (SFSB-submodel), the effect between the subject pedestrian and other pedestrians in semicircular forward space is presented in the term of the forward distance vector $\vec{L}$. The vector $\vec{I}$ (magnitude is $1/|\vec{L}|^2$ and direction is as same as that of $\vec{L}$) is chosen as input parameters of the SFSB-submodel to the learn the magnitude of the subject pedestrian's current velocity. It should be noted that the number of other pedestrians in semicircular forward space $n_c$ may be different for different subject pedestrian, causing the dimension of input parameters inconsistent. We build $n_c$ networks to solve this problem. In rectangular forward space based sub model (RFSB-submodel), pedestrian spatial distribution weighted the gauss distribution is chosen as input parameters to learn the direction of the subject pedestrian's current velocity. The networks involved in two submodels all have three layers and are trained using backpropagation algorithm. Early stop validation is applied to prevent training overfitting.

  The proposed model is applied on two typical scenarios, unidirectional flow and bidirectional flow. The simulation results are compared with experiment results from pedestrian movement trajectory and speed-density and flow-density fundamental diagrams. The results show that the simulation results are basically consistent with that in corresponding experiment. Besides, in unidirectional flow scenario, we define two evaluation indicators, mean destination error (MDE) and mean trajectory error (MTE), to quantitatively illustrate the trajectory error between simulations and experiments. $\overline{\text{MDE}}$ and $\overline{\text{MTE}}$ are calculated to be about 0.2m and 0.12m under four different densities, respectively 2.5% and 1.5% of the corridor length. In bidirectional flow scenario, relative distance error (RDE) is about 0.15m, a half of the result in [10]. And the typical lane-formation phenomena are observed in the simulation of bidirectional flow. These comparisons and analyses all indicate that the proposed model is reasonable and capable of simulating pedestrian behavior in both unidirectional and bidirectional flow scenarios illustrated in this paper.

  Although the model proposed in the paper can simulate pedestrian movement in both unidirectional and bidirectional flow scenarios in straight corridors well, there are still some problems should be noted. Firstly, hidden layers play a vital role in the performance of Back Propagation Neural Network[19]. But it is confusing that how much the number of hidden layers and number of neurons in each hidden layer should be selected [19,27–29]. Our work gave a first attempt in selection of the hidden neurons number. However, the performance of other architectures need more deeply explore. Besides, in our model, the input features of neural networks only contain the interaction between the subject pedestrian and other pedestrians or walls. While one pedestrian's movement is also relevant with his goal and original velocity (inertia factor) including magnitude and direction. These factors will be considered in the future work. In addition, it is worthy to further explore the ability of the model to describe behaviors in other scenarios like the mixed corridor scenarios with different widths.

**Acknowledgements**

This work was supported by the National Natural Science Foundation of China (Grant No. 71704168, U1933105), the Anhui Provincial Natural Science Foundation (Grant No. 1808085MG217), the Fundamental Research Funds for the Central Universities (Grant No. WK2320000040) and the State Key Laboratory of Fire Science in the University of Science and Technology of China (Grant No. HZ2018-KF12).

**References**

[1]     N.W.F. Bode, E. Ronchi, Statistical model fitting and model selection in pedestrian dynamics research, Collective Dynamics. 4 (2019) 1–32.

[2]     J. Yan-qun, Z. Peng, W. S.C., L. Ru-xun, A higher-order macroscopic model for pedestrian flows, Physica A: Statistical Mechanics and Its Applications. 389 (2010) 4623–4635. doi:10.1016/j.physa.2010.05.003.

[3]     A. Kirchner, A. Schadschneider, Simulation of evacuation processes using a bionics-inspired cellular automaton model for pedestrian dynamics, Physica A: Statistical Mechanics and Its Applications. 312 (2002) 260–276.

[4]     W. Song, X. Xu, B.-H. Wang, S. Ni, Simulation of evacuation processes using a multi-grid model for pedestrian dynamics, Physica A: Statistical Mechanics and Its Applications. 363 (2006) 492–500.

[5]     D. Helbing, I. Farkas, T. Vicsek, Simulating dynamical features of escape panic, Nature. 407 (2000) 487.

[6]     M. Chraibi, A. Seyfried, A. Schadschneider, Generalized centrifugal-force model for pedestrian dynamics, Physical Review E. 82 (2010) 46111.

[7]     L. Crociani, G. Lämmel, G. Vizzari, S. Bandini, Learning Obervables of a Multi-scale Simulation System of Urban Traffic., in: ATT@ IJCAI, 2018: pp. 40–48.

[8]     F. Martinez-Gil, M. Lozano, F. Fernández, MARL-Ped: A multi-agent reinforcement learning based framework to simulate pedestrian groups, Simulation Modelling Practice and Theory. 47 (2014) 259–275.

[9]  P. Shao, A more realistic simulation of pedestrian based on cellular automata, in: 2009 IEEE International Workshop on Open-Source Software for Scientific Computation (OSSC), IEEE, 2009: pp. 24–29.

[10] Y. Ma, E.W.M. Lee, R.K.K. Yuen, An artificial intelligence-based approach for simulating pedestrian movement, IEEE Transactions on Intelligent Transportation Systems. 17 (2016) 3159–3170. doi:10.1109/TITS.2016.2542843.

[11] X. Song, D. Han, J. Sun, Z. Zhang, A data-driven neural network approach to simulate pedestrian movement, Physica A: Statistical Mechanics and Its Applications. 509 (2018) 827–844.

[12] J. Amirian, W. Van Toll, J.-B. Hayet, J. Pettré, Data-Driven Crowd Simulation with Generative Adversarial Networks, ArXiv Preprint ArXiv:1905.09661. (2019).

[13] A. Gupta, J. Johnson, L. Fei-Fei, S. Savarese, A. Alahi, Social gan: Socially acceptable trajectories with generative adversarial networks, in: Proceedings of the IEEE Conference on Computer Vision and Pattern Recognition, 2018: pp. 2255–2264.

[14] X. Yang, L. Chen, H.H. Tian, L.Y. Dong, Experiments of unidirectional and bidirectional pedestrian flows through a bottleneck in a channel, Journal of Shanghai University. (2015).

[15] X. Ren, J. Zhang, W. Song, S. Cao, The fundamental diagrams of elderly pedestrian flow in straight corridors under different densities, Journal of Statistical Mechanics: Theory and Experiment. 2019 (2019) 23403.

[16] X. Zhou, J. Hu, X. Ji, X. Xiao, Cellular automaton simulation of pedestrian flow considering vision and multi-velocity, Physica A: Statistical Mechanics and Its Applications. 514 (2019) 982–992.

[17] J. Hu, Z. Li, H. Zhang, J. Wei, L. You, P. Chen, Experiment and simulation of the bidirectional pedestrian flow model with overtaking and herding behavior, International Journal of Modern Physics C. 26 (2015) 1550131-.

[18] C. Wu, W. Song, C. Liu, L. Lian, Modified Voronoi Method for Measuring Pedestrian Density Based on Occupy Area, Fire Safety Science. 2018, 27(2): 114-123. 27 (2018) 114–123.

[19] S. Karsoliya, Approximating number of hidden layer neurons in multiple hidden layer BPNN architecture, International Journal of Engineering Trends and Technology. 3 (2012) 714–717.

[20] Institute for Advanced Simulation 7: Civil Safety Research of Forschungszentrum Jülich, (n.d.). http://ped.fz-juelich.de/db/ (accessed June 17, 2019).

[21] M. Boltes, A. Seyfried, B. Steffen, A. Schadschneider, Automatic Extraction of Pedestrian Trajectories from Video Recordings, in: Pedestrian and Evacuation Dynamics 2008, Springer Berlin Heidelberg, Berlin, Heidelberg, 2010: pp. 43–54. doi:10.1007/978-3-642-04504-2_3.

[22] J. Zhang, W. Klingsch, A. Schadschneider, A. Seyfried, Transitions in pedestrian fundamental diagrams of straight corridors and T-junctions, Journal of Statistical Mechanics: Theory and Experiment. 2011 (2011) 1–17. doi:10.1088/1742-5468/2011/06/P06004.

[23] J. Zhang, W. Klingsch, A. Schadschneider, A. Seyfried, Ordering in bidirectional pedestrian flows and its influence on the fundamental diagram, Journal of Statistical Mechanics: Theory and Experiment. 2012 (2012). doi:10.1088/1742-5468/2012/02/P02002.

[24] S. Seer, N. Brändle, C. Ratti, Kinects and human kinetics: A new approach for studying pedestrian behavior, Transportation Research Part C: Emerging Technologies. 48 (2014) 212–228.

[25] J. Ma, W. Song, Z. Fang, S. Lo, G. Liao, Experimental study on microscopic moving characteristics of pedestrians in built corridor based on digital image processing, Building and Environment. 45 (2010) 2160–2169.

[26] M. Zhang, H. Yuan, The PauTa criterion and rejecting the abnormal value, Journal of Zhengzhou University of Technology. 1 (1997) 84–88.

[27] J. Ke, X. Liu, Empirical analysis of optimal hidden neurons in neural network modeling for stock prediction, in: 2008 IEEE Pacific-Asia Workshop on Computational Intelligence and Industrial Application, IEEE, 2008: pp. 828–832.

[28] D. Hunter, H. Yu, M.S. Pukish III, J. Kolbusz, B.M. Wilamowski, Selection of proper neural network sizes and architectures—A comparative study, IEEE Transactions on Industrial Informatics. 8 (2012) 228–240.

[29] G.-B. Huang, Learning capability and storage capacity of two-hidden-layer feedforward networks, IEEE Transactions on Neural Networks. 14 (2003) 274–281.